# Low-Frequency Noise in Quasi-1D (TaSe$_4$)$_2$I Weyl Semimetal Nanoribbons

Subhajit Ghosh[1,2], Fariborz Kargar[1,2], Nick R. Sesing[3], Zahra Barani[1,2], Tina T. Salguero[3], Dong Yan[4], Sergey Rumyantsev[5], and Alexander A. Balandin[1,2*]

[1]Nano-Device Laboratory, Department of Electrical and Computer Engineering, Bourns College of Engineering, University of California, Riverside, California 92521 U.S.A.

[2]Phonon Optimized Engineered Materials Center, Bourns College of Engineering, University of California, Riverside, California 92521 U.S.A.

[3]Department of Chemistry, University of Georgia, Athens, Georgia 30602 U.S.A.

[4]Center for Nanoscale Science and Engineering, University of California, Riverside, California 92521 U.S.A.

[5]CENTERA Laboratories, Institute of High-Pressure Physics, Polish Academy of Sciences, Warsaw 01-142 Poland

---

[*] Corresponding author (A.A.B.): balandin@ece.ucr.edu ; web-site: http://balandingroup.ucr.edu/

## Abstract

We report on low-frequency current fluctuations, *i.e.* electronic noise, in quasi-one-dimensional (TaSe$_4$)$_2$I Weyl semimetal nanoribbons. It was found that the noise spectral density is of the 1/*f* type and scales with the square of the current, $S_I \sim I^2$ (*f* is the frequency). The noise spectral density increases by almost an order of magnitude and develops Lorentzian features near the temperature $T \sim 225$ K. These spectral changes were attributed to the charge-density-wave phase transition even though the temperature of the noise maximum deviates from the reported Peierls transition temperature in bulk (TaSe$_4$)$_2$I crystals. The noise level, normalized by the channel area, in these Weyl semimetal nanoribbons was surprisingly low, $\sim 10^{-9}$ μm$^2$Hz$^{-1}$ at *f*=10 Hz, when measured below and above the Peierls transition temperature. Obtained results shed light on the specifics of electron transport in quasi-1D topological Weyl semimetals and can be important for their proposed applications as downscaled interconnects.

**1. Introduction**

Recently, one-dimensional (1D) van der Waals (vdW) quantum materials with emergent topological phases, derived from strongly correlated interactions, have attracted significant attention [1, 3]. A prototype example is a tetra-selenide compound (TaSe$_4$)$_2$I, with a structure featuring unusual axially chiral (TaSe$_4$)$_n$ chains [4-7]. At elevated temperatures, this quasi-1D material is defined as a Weyl semimetal with Weyl points located above and below the Fermi level, forming pairs with the opposite chiral charge. At temperatures below the Peierls transition temperature $T_P$ = 248 K – 263 K, (TaSe$_4$)$_2$I reveals the charge-density-wave (CDW) phase [4, 8-13]. The quantum CDW phase consists of a periodic modulation of the electronic charge density accompanied by a periodic distortion of the atomic lattice, in this case, Ta tetramerization [14-19]. It has been suggested that (TaSe$_4$)$_2$I reveals a correlated topological phase, which arises from the formation of CDW in a Weyl semimetal [11, 12]. This quasi-1D quantum material represents an interesting avenue for exploring the interplay of correlations and topology, as well as being an exciting system for examining new functionalities and electronic applications [17, 20].

There is an interesting applied physics aspect in the research of topological semimetals. The resistance and current density bottlenecks in downscaled metal interconnect motivate the search for new materials for the back end of the line (BEOL) interconnect applications [21, 22]. When the interconnect linewidth scales below the electron mean free path, its resistivity increases in a power-law function due to increased electron scattering from interfaces and grain boundaries [23, 24]. This problem is persistent for all *elemental* metals, including Cu, Co, and Ru. Quasi-1D vdW materials demonstrated different dependencies where the resistivity remained nearly constant with the scaling of the interconnect cross-sectional area due to their single-crystal nature and sharp vdW boundary interfaces [25, 26]. There are indications that in some topological semimetals, the electrical conductivity can actually increase as the cross-sectional area decreases. Recent studies have shown that in the topological Weyl semimetal NbAs the resistivity can decrease by an order of magnitude from 35 µΩ-cm in bulk single crystals to 1–5 µΩ-cm in ~200 nm nanoribbons [27]. For this reason, a better understanding of electron transport phenomena in Weyl semimetals can have an immediate practical significance for interconnect applications [28, 29].

Here, we report on the low-frequency current fluctuations, *i.e.*, low-frequency electronic noise, also referred to as *excess* noise, in (TaSe$_4$)$_2$I nanoribbons. The low-frequency noise includes the 1/$f$ and generation-recombination (G-R) noise with a Lorentzian type spectrum, which comes on top of the thermal and shot noise background ($f$ is the frequency). It is known that 1/$f$ noise can provide information on the electron transport and charge carrier recombination in a given material, as well as serve as an early indicator of electromigration damage [30-35]. We have previously used low-frequency noise measurements for monitoring phase transition in various materials [36-39] as well as for assessing the material quality and device reliability [40-42]. In this work, we are primarily motivated by the following questions. Is electron transport in the topological Weyl semimetals characterized by inherently lower noise owing to the suppression of certain electron scattering channels? Can one use the excess noise data to verify the CDW transitions in the topological Weyl semimetals? We are also interested in assessing if Weyl semimetals are acceptable for interconnect applications in terms of their electronic noise level.

## 2. Experimental Details

The crystal structure of quasi-1D (TaSe$_4$)$_2$I above $T_P$ is illustrated in Figure 1 (a) with a view showing (TaSe$_4$)$_n$ chains aligned along the *c*-axis, within an iodide lattice [43]. The view of a single (TaSe$_4$)$_n$ chain highlights several key features, including the coordination of each Ta center to eight selenium atoms in rectangular anti-prismatic geometry, the equidistant Ta centers at ~3.2 Å that support metallic bonding, the slightly asymmetric bridging of Se$_2^{2-}$ pairs between Ta centers, in part due to interactions with iodide, and most notably, the rotating pattern of Se$_2^{2-}$ about the chain axis that generates axial chirality. Single crystal (TaSe$_4$)$_2$I source materials were synthesized by the chemical vapor transport (CVT) method from stoichiometric amounts of tantalum and selenium, and an excess of iodine that serves as both reactant and transport agent. Using a temperature gradient of 590–530 °C for 10 d provided mm- to cm-sized (TaSe$_4$)$_2$I crystals in good yield [44, 45]. The quality of this material was examined using several methods (see Figure 1 (b-c) and the Supplemental Materials). Scanning electron microscopy (SEM) reveals features exhibited by many 1D van der Waals materials, such as growth striations and facile cleavage along the van der Waals gap. Powder x-ray diffraction (XRD) confirms the expected structure of (TaSe$_4$)$_2$I. Energy dispersive x-ray spectroscopy (EDS) demonstrates homogeneity and provides a composition of (TaSe$_{3.4}$)$_2$I$_{0.8}$; selenium and iodide deficiencies are well-known for metal chalcogenide compounds and expected here [46, 47].

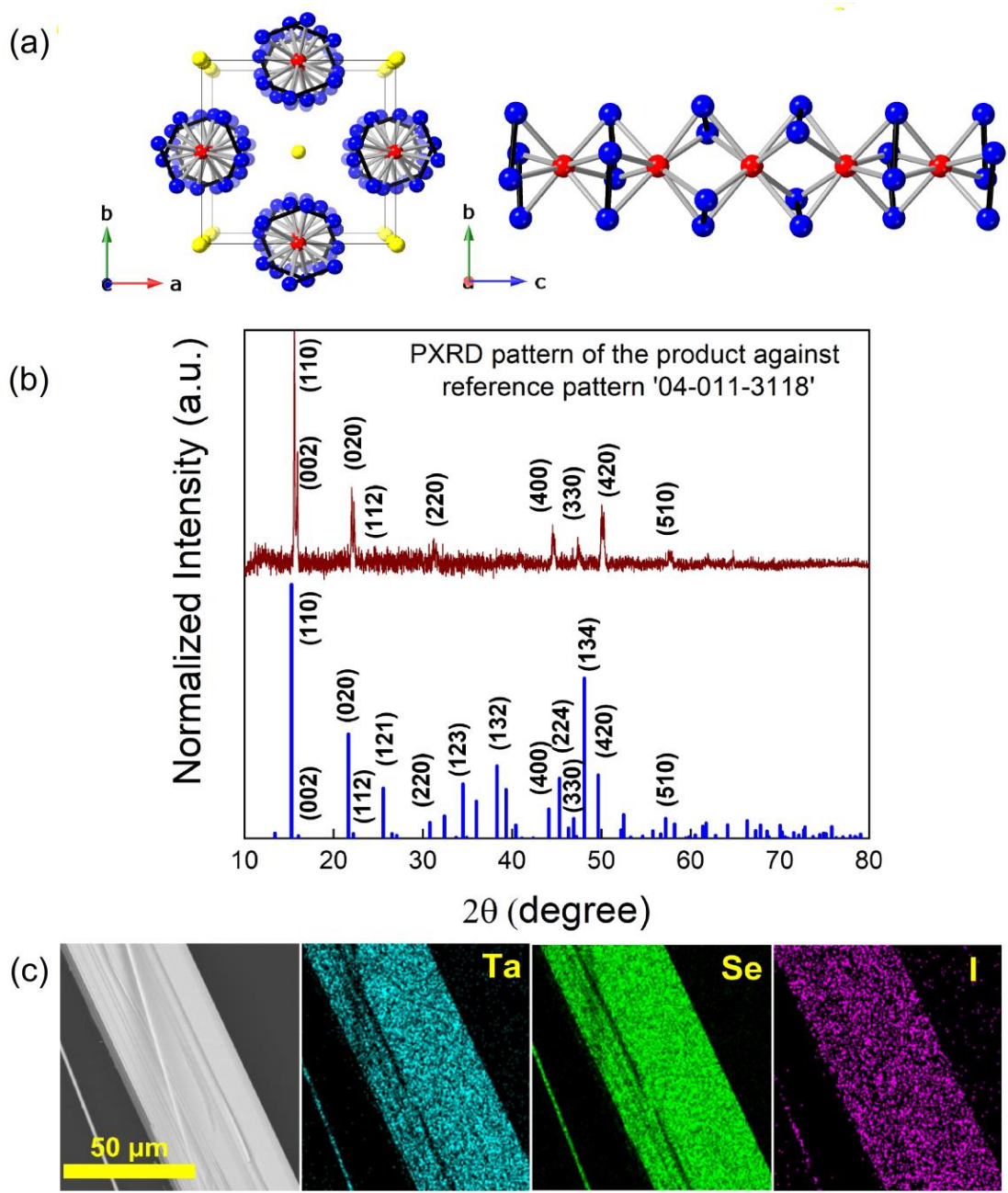

**Figure 1:** (a) Crystal structure of (TaSe$_4$)$_2$I viewed down the *c*-axis (left panel) and *a*-axis (right panel), with atoms corresponding to Ta (red), Se (blue), and I (yellow). (b) Powder x-ray diffraction pattern of the CVT-grown crystals; experimental (top) and reference card 04-011-3118 (bottom pattern). (c) Scanning electron microscopy image of a mechanically exfoliated (TaSe$_4$)$_2$I crystal surface and corresponding energy dispersive spectroscopy elemental mapping.

The quasi-1D nanoribbons of (TaSe$_4$)$_2$I were prepared using a mechanical exfoliation technique on top of clean Si/SiO$_2$ substrates (University Wafer, *p*-type Si/SiO$_2$, <100>). Here, we use the term nanoribbon rather than nanowire to describe these structures owing to the fact that the width of the selected structures was substantially wider than the thickness. This allowed for more accurate nanofabrication, testing, and comparison with quasi-2D materials. The exfoliated (TaSe$_4$)$_2$I nanoribbons had a length of a few micrometers, a width on the scale of hundred nanometers, and a thickness in the range of 10 nm – 100 nm as confirmed by atomic force microscopy (AFM). The AFM images for two exfoliated nanoribbons with different thicknesses are shown in Figure 2 (a). A representative Raman spectrum of the exfoliated nanoribbons is provided in Figure 2 (b). One can clearly see seven Raman peaks at the frequencies of 63.9 cm$^{-1}$, 67.9 cm$^{-1}$, 100.1 cm$^{-1}$, 143.3 cm$^{-1}$, 160.3 cm$^{-1}$, 182.8 cm$^{-1}$, and 270.9 cm$^{-1}$, in line with prior reports on bulk (TaSe$_4$)$_2$I crystals [48-51]. All observed Raman frequencies belong to the A$_1$ vibrational mode type, except the peak at 67.9 cm$^{-1}$ which belongs to the B$_2$ symmetry group [49]. The test structures with multiple electrodes were prepared using electron-beam lithography (EBL) to define the contacts on the same nanoribbon. Electron beam evaporation (EBE) was used to deposit Cr/Au metals (10 nm / 100 nm) to form the contacts for measurements (refer to the Supplements for detailed fabrication steps). Figure 2 (c) shows a schematic diagram of the test structure, containing several metal contacts and pads on the Si/SiO$_2$ substrate. The channel lengths of the individual devices, *i.e.*, the distance between two contacts, were in the range of 1 μm to 6 μm. The quality of the contacts and the nanoribbon channels was verified with SEM. A colored SEM image of a representative (TaSe$_4$)$_2$I nanoribbon test structure is provided in Figure 2 (d).

### 3. Results and Discussions

The temperature-dependent current-voltage (I-V) measurements of the fabricated nanoribbon test structures were carried out inside a cryogenic probe station (Lakeshore TTPX) under vacuum using a semiconductor analyzer (Agilent B1500). The low-frequency noise measurements were performed using an in-house built system. The noise measurement circuit consists of a low-noise DC battery, a potentiometer (POT), and a load resistor connected in series to the device under test (DUT) kept inside the probe station chamber. The POT controls the voltage drops between the load and the DUT of the voltage divider noise circuit. The load resistor was kept grounded in this configuration. During the noise measurements, the voltage fluctuations at the output were transferred to a low noise preamplifier (SR-560) which amplified the signal and sent it to a signal analyzer. The signal analyzer transformed the time domain signal to its corresponding frequency domain. In our noise calculations, the voltage spectral density, $S_V$, was recalculated to its equivalent current spectral density, $S_I$, and normalized by the corresponding current squared, $I^2$. Further details of our noise measurement systems and procedures can be found in the Supplements and in the prior reports for other materials and devices [39, 40, 52, 53].

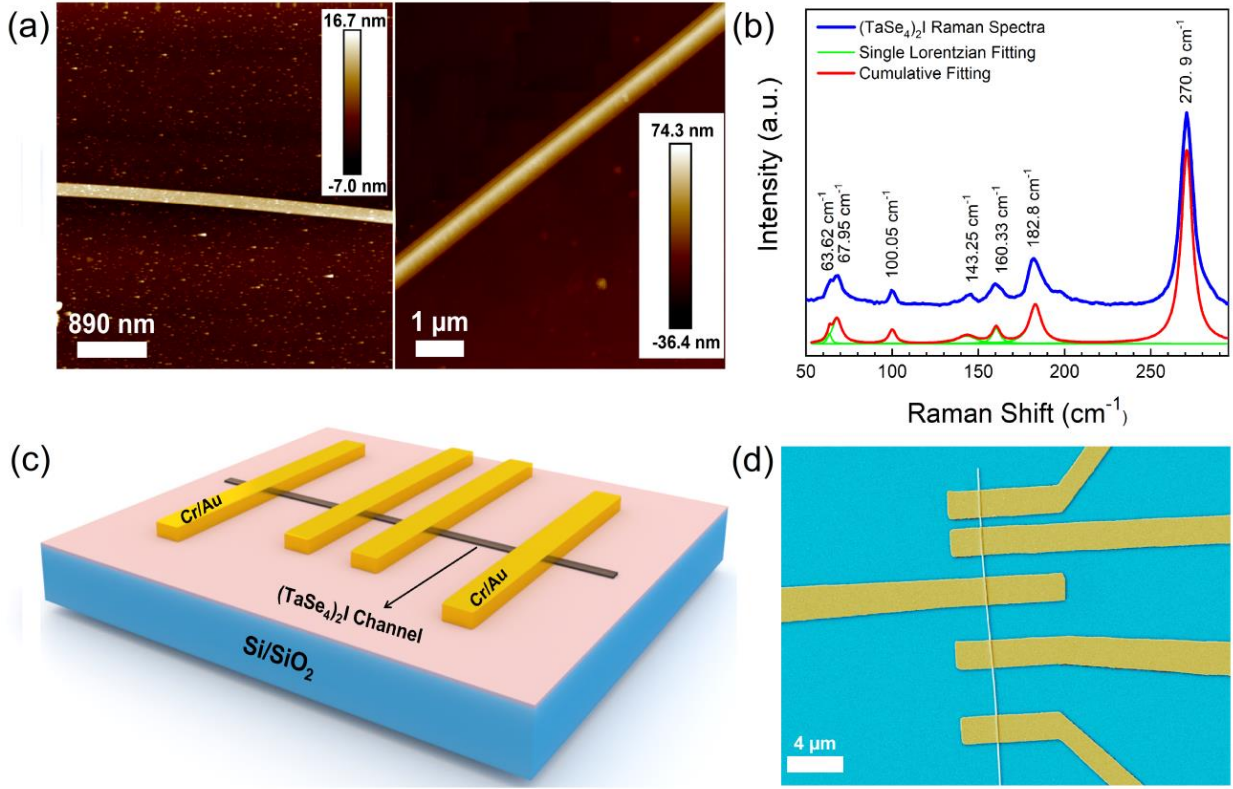

**Figure 2:** Characterization of the exfoliated quasi-one-dimensional (TaSe$_4$)$_2$I nanoribbons. (a) Atomic force microscopy images of two exfoliated nanoribbons with different thicknesses and widths. The studied nanoribbons had thicknesses in the range of 10 nm to 100 nm. (b) Raman spectrum of a (TaSe$_4$)$_2$I nanoribbon at room temperature. (c) Schematic of a (TaSe$_4$)$_2$I nanoribbon test structure on Si/SiO$_2$ substrate. (d) Scanning electron microscopy image of a (TaSe$_4$)$_2$I nanoribbon test structure with varying channel lengths from 1 µm to 4 µm. Pseudo-colors are used for clarity.

To study the electrical characteristics of low-dimensional materials, which reveal phase transitions, it is important to verify the quality of the electrical contacts. Figure 3 (a) presents the low-bias I-V characteristics of one of the (TaSe$_4$)$_2$I test structures for several channel lengths, *i.e.*, I-Vs measured between different pairs of contacts. The I-Vs show linear behavior across the measured bias ranges confirming high-quality Ohmic contacts. The contact resistance of (TaSe$_4$)$_2$I nanoribbon was determined using the conventional transmission line measurement (TLM) technique (see Figure 3 (b). The contact resistance value for this representative device is $2R_C$=440 Ω, which is an order of magnitude lower than any of the channel resistances, *R*. The latter further confirms the quality of the fabricated contacts. The fact that $R_C<<R$ is beneficial for the interpretation of the noise measurements as well.

The as-measured low-frequency voltage noise spectral density, $S_V$, is shown in Figure 3 I. The room-temperature data are presented for a (TaSe$_4$)$_2$I nanoribbon device with the 2-µm channel length measured for the source-drain bias, $V_D$, ranging from 10 mV to 60 mV. The noise spectra for all bias voltages are of $1/f^{\gamma}$ ($\gamma \approx 1$) flicker noise type, which is typical for both semiconductor and metallic materials [30, 38, 39, 54]. The noise level measured at the lowest bias point is at least an order of magnitude higher than the background noise of the measurement system, confirming that the flicker $1/f$ noise is intrinsic to the DUT and not from any other source. The corresponding current spectral density, $S_I$, as a function of the source-drain current, *I*, at a fixed frequency of *f*=10 Hz is presented in Figure 3 (d). The $S_I$ *vs. I* behavior is quadratic, *i.e.*, $S_I \sim I^2$, with the exact slope of 1.98. The quadratic scaling of the noise spectral density, $S_I$, with the source-drain current is expected for any linear resistor. Thus, we verified the accuracy of the noise measurement procedures and the fact that the (TaSe$_4$)$_2$I nanoribbons act as passive linear resistors.

The temperature-dependent I-Vs and noise data for a (TaSe$_4$)$_2$I device with a 2-µm channel length are presented in Figure 4 (a-d) (refer to Figure S3 for the I-V data of a different (TaSe$_4$)$_2$I nanoribbon device). Figure 4 (a) shows the device resistance in a logarithmic scale normalized

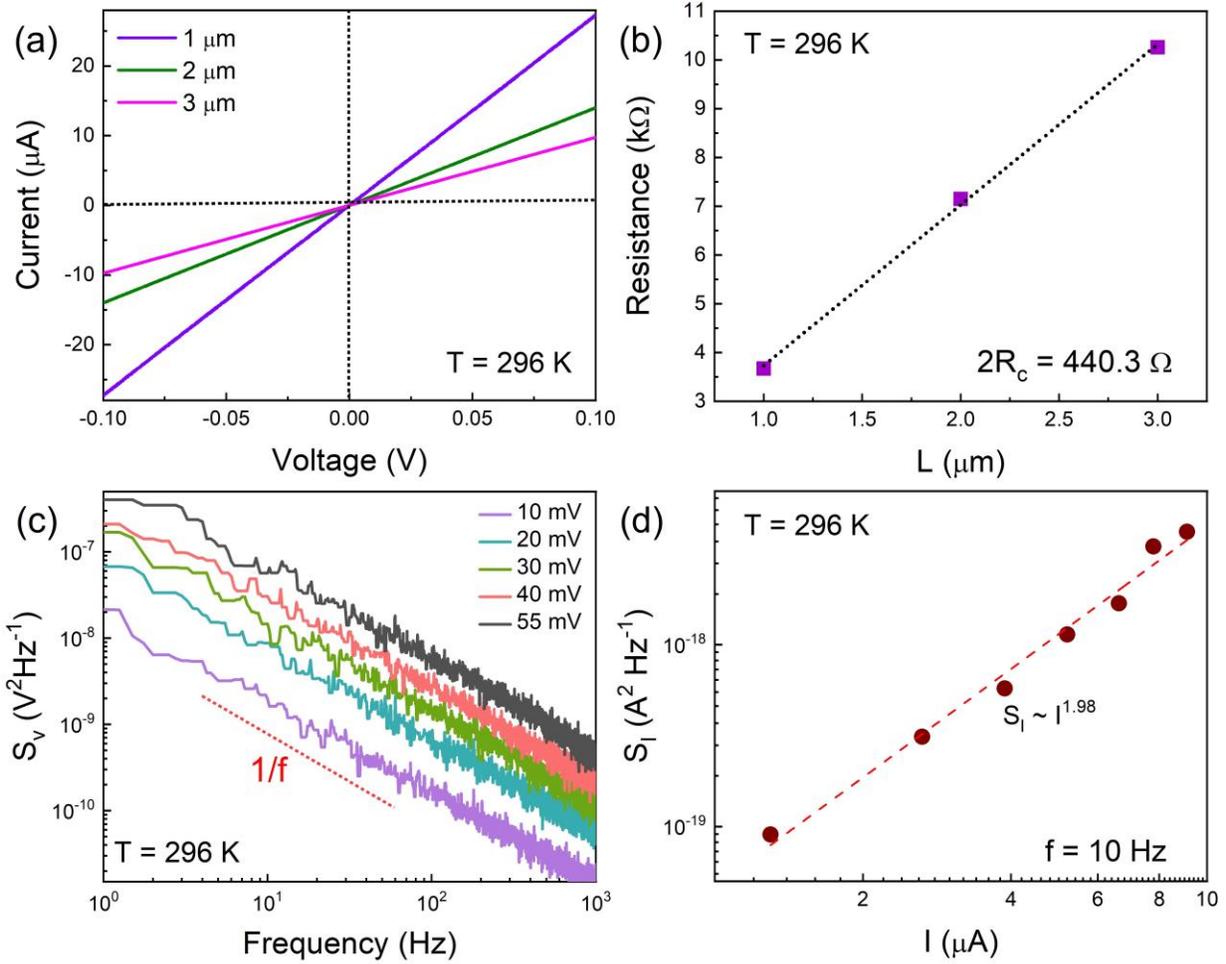

**Figure 3:** Electrical and low-frequency noise characteristics of a (TaSe$_4$)$_2$I nanoribbons at room temperature. (a) Current-voltage characteristics for nanoribbon devices with different channel lengths. (b) Resistance of the nanoribbon devices as a function of the channel lengths. (c) Voltage noise spectral density, $S_V$, as a function of frequency for a (TaSe$_4$)$_2$I nanoribbon device at different source-drain biases. (d) Current noise spectral density, $S_I$, as a function of the device current.

by the resistance of the channel at 300 K, *i.e. log[R/R$_{300}$]*, as a function of the inverse temperature, *i.e.* $10^3/T$. The I-V measurements were conducted both in the heating and cooling cycles. Overall, the dependence of the *log[R/R$_{300}$]* on inverse temperature is consistent with prior reports for bulk (TaSe$_4$)$_2$I samples [4, 5, 8-13]. There is an abrupt change in the slope of the resistance below RT. This slope change is more clearly observed in the plot of the derivative characteristics presented in Figure 4 (b). In our case, we observed the transition at *T*=235 K for both cooling and heating cycle measurements. Previous reports attributed the change in the resistivity slope to the Peierls transition, *i.e.*, CDW phase transition, observed mostly at *T$_P$*=260

- 263 K [4, 8, 9]. However, some reports indicated this transition at a temperature as low as $T_P$=235 - 240 K [55, 56]. It is known from experience with other CDW materials, that the temperature of the CDW phase transition may depend on the sample thickness [57, 58]. In addition, some data scatter for the transition temperature can be due to small stoichiometric variations, *e.g.*, the loss of iodine.

It has been stated that the Peierls transition in (TaSe$_4$)$_2$I is accompanied by opening a CDW energy bandgap of ~0.2 eV [4, 12, 59, 60]. There is an unusual feature of the phase transition in (TaSe$_4$)$_2$I, which was noticed and discussed in the original studies of bulk crystals [8-10]. The material reveals a non-metallic *R(T)* dependence both below and *above* $T_P$. This issue was addressed in detail in a report that described (TaSe$_4$)$_2$I as the zero-bandgap semiconductor and introduced a notion of the semiconductor-semiconductor phase transition [59]. In our measurements with (TaSe$_4$)$_2$I nanoribbons, the resistance change near $T_P$ is consistently observed but it is somewhat more gradual than in the case of bulk samples. We speculate that this can be related to the strain induced by the lattice mismatch between the material and Si/SiO$_2$ substrate. The latter is supported by our experiments with Al$_2$O$_3$ and other substrates and prior reports on the effect of the substrate-induced strain on resistive switching in nanowires [13, 61 - 63]. Based on the above considerations, we can conclude that our measured resistivity data are in line with previous reports [8-11, 13], and focus on the current fluctuations in (TaSe$_4$)$_2$I nanoribbons.

Figure 4 (c) shows the normalized noise current spectral density, $S_I/I^2$, as a function of frequency, *f*, at temperatures near $T_P$ for the same device. The noise measurements were conducted between 200 K – 235 K to capture the Peierls transition from a possible change in the noise spectrum. One can notice an evolution of the 1/*f* spectrum to Lorentzian bulge near the Peierls transition temperature. To clarify the trend, we plotted $f \times S_I/I^2$ *vs. f*, which removes the 1/*f* background. One can see that at *T*=225 K, the noise spectral density reaches a maximum and develops a Lorentzian bulge. As the temperature increases further the noise becomes 1/*f* again, and its level decreases. We argue that the noise increase and Lorentzian feature are signatures of the Peierls transition, which we observed in the resistivity behavior in Figures 4 (a) and (b). The same trend – noise increases and Lorentzian – type bulges near the CDW phase transitions have been reported previously for different materials [36-39]. Some differences in temperature $T_P$ extracted from the resistivity and noise data can be explained by a difference in the rate at which the temperature was changed in the probe station during these two independent measurements. In addition, there is a possibility of a temperature drift during the noise measurements. Generally, the Lorentzian noise spectrum is a signature of a two-level system [64, 65]. In the case of a phase transition, the material state and its resistance can switch between the two phases until the material system is driven further away from the transition point $T_P$. We observed a similar behavior in 1T-TaS$_2$, another CDW material [36, 37]. It is unlikely that the Lorentzian bulges which we see in the noise spectrum of (TaSe$_4$)$_2$I are due to the generation-recombination (G-R) noise that originates from high concentrations of one type of defects with particular time constants since it is observed only at one temperature and the noise spectrum returns to its original 1/*f* type.

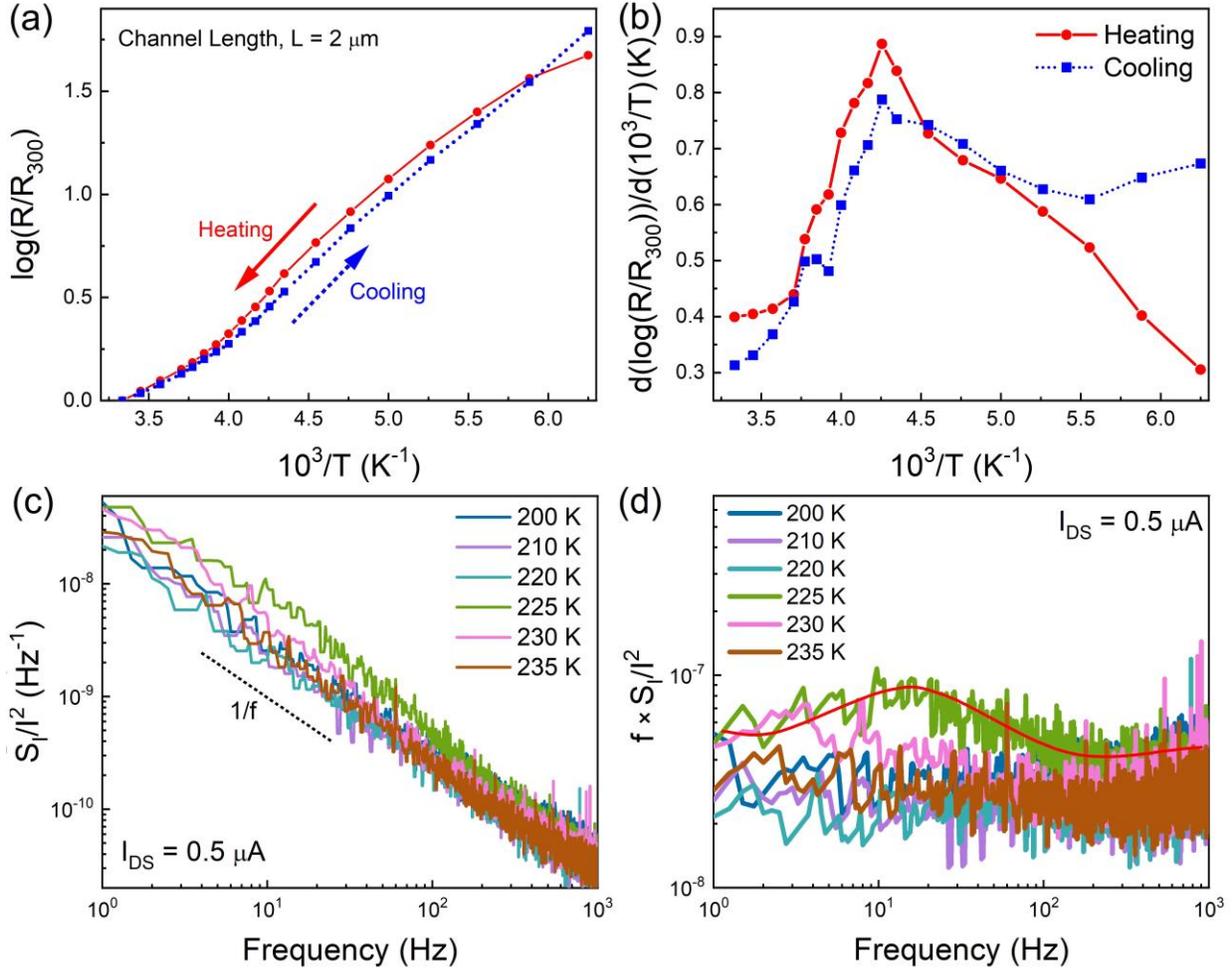

**Figure 4:** Temperature-dependent electrical and low-frequency noise characteristics of a (TaSe$_4$)$_2$I nanoribbon device. (a) Logarithmic normalized resistance, $log([R/R_{300}]$, as a function of inverse temperature, $10^3/T$, for a (TaSe$_4$)$_2$I nanoribbon device with a 2-µm channel length. (b) Logarithmic derivative, $d(log([R/R_{300}])/d(10^3/T)$, vs inverse temperature of the same device. (c) Normalized noise current spectral density, $S_I/I^2$, as a function of frequency at temperatures near the transition temperature measured at a constant device current of 0.5 µA. The noise behavior is of *1/f* type, except at $T \sim 225$ K, where the noise becomes Lorentzian type. (d) Normalized noise spectral density multiplied by the frequency, $f \times S_I/I^2$, as a function of frequency at different temperatures.

In Figure 5, we plot the noise spectral density normalized for the channel area, $\beta=S_I/I^2\times(W\times L)$, at a fixed frequency *f*=10 Hz and two representative values of the current. We have previously introduced the $\beta$ parameter in order to compare low-frequency noise levels in two-dimensional (2D) materials such as graphene and MoS$_2$ [66, 67]. One can see that the noise spectral density increases by an order of magnitude near $T_P$ proving that the noise level is a suitable indicator of the phase transition. Figure 5 attests to the low noise level, normalized by the channel area, compared to 2D materials. The nanoribbon shape of the (TaSe$_4$)$_2$I channels, with the width much larger than the thickness, makes the comparison with 2D materials meaningful. The value of the $\beta$ parameter away from the phase transition is below $2 \times 10^{-9}$ µm$^2$Hz$^{-1}$. For comparison, the area-normalized noise level in graphene is $\beta=10^{-8}$ µm$^2$Hz$^{-1}$ while that in thin MoS$_2$ is $\beta=10^{-5}$ µm$^2$Hz$^{-1}$ [60]. The noise level, $S_I/I^2$, without surface normalization, is also rather low, below $4 \times 10^{-9}$ Hz$^{-1}$, away from the phase transition point, both in the low and high-temperature regions.

It is not possible to state at this point if the low level of the low-frequency noise in (TaSe$_4$)$_2$I nanoribbons is due to the current fluctuation suppression in the topologically protected conductive channels of the Weyl semimetal [68, 69]. The latter would require a dedicated theoretical study allowing to connect the electron scattering and capture rates with the fluctuations in the number of electrons and their mobility. The experimental data reported in this study indicates an unexpectedly low noise level and motivates future studies. One should also note that the noise level measured in this Weyl semimetal nanoribbons is sufficiently low for interconnect applications. The resistivity extracted for these nanoribbons was similar to the values reported for bulk (TaSe$_4$)$_2$I, on the order of $10^{-3}$ Ω-cm [9, 11]. This material may not be the optimum one for interconnect applications in terms of its resistivity, but the low noise level for topological Weyl semimetals is a promising feature. It is known that low-frequency noise can be an early indicator of damage to materials and devices [70-73]. With the device degradation, the noise increases at a much faster rate than the changes in the averaged characteristics such as I-Vs. The latter makes the noise a sensitive predictor of a lifetime. The results obtained in this work can be used for developing assessment methodologies for the reliability of topological semimetals.

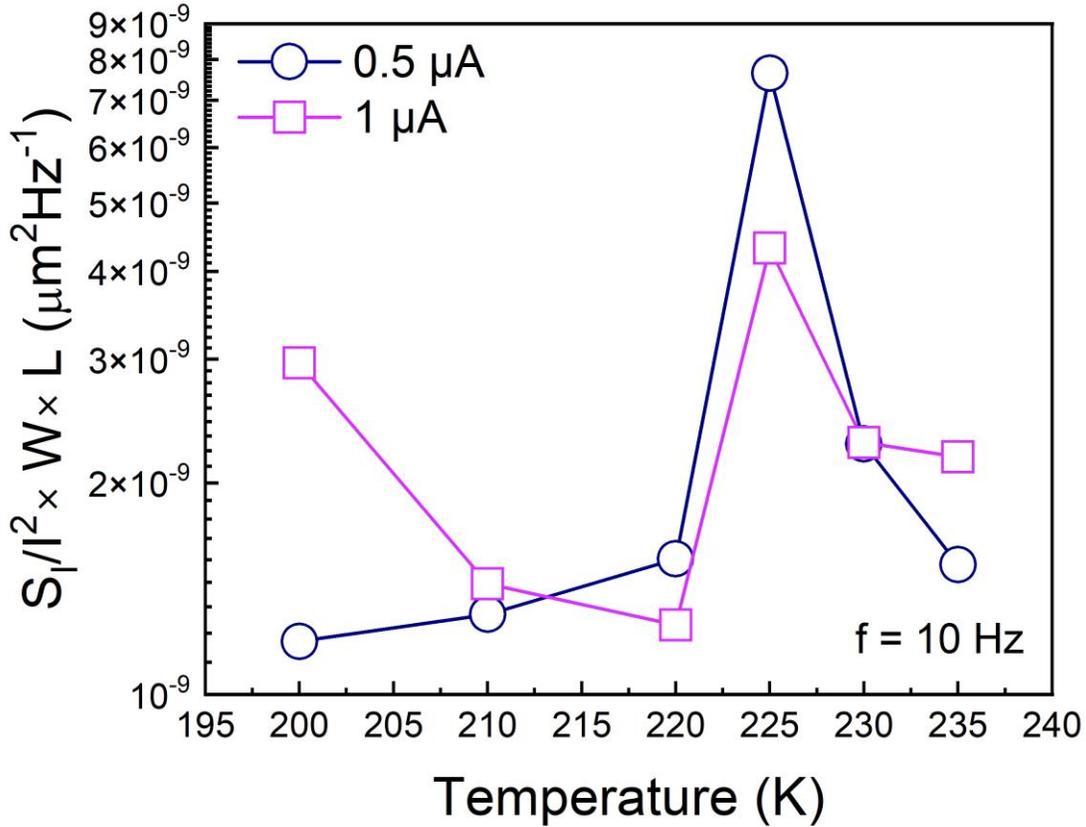

**Figure 5:** Evolution of the noise spectral density with temperature. The noise spectral density normalized by device current and channel area, $S_I/I^2 \times W \times L$, measured at $f = 10$ Hz, at the constant currents of $I_{DS} = 0.5$ µA and 1.0 µA. The noise level increases substantially near $T \sim 225$ K.

## 4. Conclusions

In conclusion, we describe low-frequency current fluctuations in quasi-1D (TaSe$_4$)$_2$I Weyl semimetal nanoribbons. The noise spectral density increases by almost an order of magnitude and develops Lorentzian features near the temperature *T*~225 K. These spectral changes were attributed to the CDW phase transition even though the temperature of the noise maximum deviates from the reported Peierls transition temperature in bulk crystals. The noise level normalized to the device area in the Weyl semimetal nanowires is surprisingly low, $S_I/I^2 \times (W \times L)$ ~$10^{-9}$ Hz$^{-1}$ at *f*=10 Hz, when measured below and above the Peierls transition temperature. This value is an order of magnitude lower than that in graphene and other quasi-2D materials. These results shed light on the specifics of electron transport in quasi-1D topological Weyl semimetals and can be important for their proposed applications as downscaled interconnects.

**Supporting Information**

Supporting Information is available from the Wiley Online Library or from the author.

**Acknowledgments**

A.A.B. was supported by the Vannevar Bush Faculty Fellowship from the Office of Secretary of Defense (OSD) under the Office of Naval Research (ONR) contract N00014-21-1-2947. The work of T.T.S. and N.R.S. was supported by the subcontract of this ONR award. A.A.B. also acknowledges the support from the National Science Foundation (NSF) program Designing Materials to Revolutionize and Engineer our Future (DMREF) *via* a project DMR-1921958 entitled Collaborative Research: Data-Driven Discovery of Synthesis Pathways and Distinguishing Electronic Phenomena of 1D van der Waals Bonded Solids. S.R. was supported, in part, by the International Research Agendas program of the Foundation for Polish Science co-financed by the European Union under the European Regional Development Fund (No. MAB/2018/9).

**Author Contributions**

A.A.B. conceived the idea, coordinated the project, and led the experimental data analysis and manuscript preparation. S.G. fabricated the test structures, performed Raman spectroscopy, measured I-Vs, and electronic noise characteristics, and contributed to data analysis. N.R.S. synthesized the bulk crystals and conducted material characterization. T.T.S. supervised material synthesis and contributed to material data analysis. Z.B. assisted with Raman spectroscopy and contributed to material characterization. D.Y. assisted with cleanroom nanofabrication. S.R. and F.K. contributed to the noise data analysis. All authors contributed to the manuscript preparation.

**Conflict of Interest**

The authors declare no conflict of interest.

**The Data Availability Statement**

The data that support the findings of this study are available from the corresponding author upon reasonable request.

**REFERENCES**

1. Y. Yuan, W. Wang, Y. Zhou, X. Chen, C. Gu, C. An, Y. Zhou, B. Zhang, C. Chen, R. Zhang, Z. Yang, *Advanced Electronic Materials* **2020**, *6*, 1901260.
2. O. F. Shoron, M. Goyal, B. Guo, D. A. Kealhofer, T. Schumann, S. Stemmer, *Advanced Electronic Materials* **2020**, *6*, 2000676.
3. I. D. Bernardo, J. Hellerstedt, C. Liu, G. Akhgar, W. Wu, S. A. Yang, D. Culcer, S. K. Mo, S. Adam, M. T. Edmonds, M. S. Fuhrer, *Advanced Materials* **2021**, *33*, 2005897.
4. J. Gooth, B. Bradlyn, S. Honnali, C. Schindler, N. Kumar, J. Noky, Y. Qi, C. Shekhar, Y. Sun, Z. Wang, B. A. Bernevig, C. Felser, *Nature* **2019**, *575*, 315.
5. A. A. Sinchenko, R. Ballou, J. E. Lorenzo, T. Grenet, P. Monceau, *Applied Physics Letters* **2022**, *120*, 063102.
6. W. Shi, B. J. Wieder, H. L. Meyerheim, Y. Sun, Y. Zhang, Y. Li, L. Shen, Y. Qi, L. Yang, J. Jena, P. Werner, K. Koepernik, S. Parkin, Y. Chen, C. Felser, B. A. Bernevig, Z. Wang, *Nature Physics* **2021**, *17*, 381.
7. T. Konstantinova, L. Wu, W. G. Yin, J. Tao, G. D. Gu, X. J. Wang, J. Yang, I. A. Zaliznyak, Y. Zhu, *npj Quantum Materials* **2020**, *5*, 1.
8. M. Maki, M. Kaiser, A. Zettl, G. Grüner, *Solid State Communications* **1983**, *46*, 497.
9. Z. Z. Wang, M. C. Saint-Lager, P. Monceau, M. Renard, P. Gressier, A. Meerschaut, L. Guemas, J. Rouxel, *Solid State Communications* **1983**, *46*, 325.
10. H. Fujishita, M. Sato, S. Hoshino, *Solid State Communications* **1984**, *49*, 313.
11. I. A. Cohn, S. G. Zybtsev, A. P. Orlov, S. v. Zaitsev-Zotov, *JETP Letters* **2020**, *112*, 88.
12. X. P. Li, K. Deng, B. Fu, Y. K. Li, D. S. Ma, J. F. Han, J. Zhou, S. Zhou, Y. Yao, *Physical Review B* **2021**, *103*, L081402.
13. Q. G. Mu, D. Nenno, Y. P. Qi, F. R. Fan, C. Pei, M. Elghazali, J. Gooth, C. Felser, P. Narang, S. Medvedev, *Physical Review Materials* **2021**, *5*, 084201.
14. G. Grüner, *Reviews of Modern Physics* **1988**, *60*, 1129.
15. S. v Zaitsev-Zotov, *Physics-Uspekhi* **2004**, *47*, 533.
16. P. Monceau, *Advances in Physics* **2012**, *61*, 325.
17. A. A. Balandin, S. v. Zaitsev-Zotov, G. Grüner, *Applied Physics Letters* **2021**, *119*, 170401.

18. V. Favre-Nicolin, S. Bos, J. E. Lorenzo, J.-L. Hodeau, J.-F. Berar, P. Monceau, R. Currat, F. Levy, H. Berger, *APS* 2001, *87*, 015502.

19. Y. Shi, Q. Li, J. Yu, al -, S. Onishi, M. Jamei, A. Zettl -, S. van Smaalen, E. J. Lam, *Journal of Physics: Condensed Matter* 2001, *13*, 9923.

20. A. A. Balandin, F. Kargar, T. T. Salguero, R. K. Lake, *Materials Today* **2022**, *55*, 74.

21. Z. Chen, N. Boyajian, Z. Lin, R. T. Yin, S. N. Obaid, J. Tian, J. A. Brennan, S. W. Chen, A. N. Miniovich, L. Lin, Y. Qi, X. Liu, I. R. Efimov, L. Lu, *Advanced Materials Technologies* **2021**, *6*, 2100225.

22. A. Ruiz-Clavijo, O. Caballero-Calero, D. Navas, A. A. Ordoñez-Cencerrado, J. Blanco-Portals, F. Peiró, R. Sanz, M. Martín-González, *Advanced Electronic Materials* **2022**, 2200342.

23. W. Steinhögl, G. Schindler, G. Steinlesberger, M. Engelhardt, *Physical Review B* **2002**, *66*, 075414.

24. D. Josell, S. H. Brongersma, Z. Tokei, *Annual Review of Materials* **2009**, *39*, 231.

25. M. A. Stolyarov, G. Liu, M. A. Bloodgood, E. Aytan, C. Jiang, R. Samnakay, T. T. Salguero, D. L. Nika, S. L. Rumyantsev, M. S. Shur, K. N. Bozhilov, A. A. Balandin, *Nanoscale* **2016**, *8*, 15774.

26. T. A. Empante, A. Martinez, M. Wurch, Y. Zhu, A. K. Geremew, K. Yamaguchi, M. Isarraraz, S. Rumyantsev, E. J. Reed, A. A. Balandin, L. Bartels, *Nano Letters* **2019**, *19*, 4355.

27. C. Zhang, Z. Ni, J. Zhang, X. Yuan, Y. Liu, Y. Zou, Z. Liao, Y. Du, A. Narayan, H. Zhang, T. Gu, X. Zhu, L. Pi, S. Sanvito, X. Han, J. Zou, Y. Shi, X. Wan, S. Y. Savrasov, F. Xiu, *Nature Materials* **2019**, *18*, 482.

28. C. T. Chen, U. Bajpai, N. A. Lanzillo, C. H. Hsu, H. Lin, G. Liang, *Technical Digest - International Electron Devices Meeting, IEDM* **2020**, 32.4.1.

29. D. Gall, J. J. Cha, Z. Chen, H. J. Han, C. Hinkle, J. A. Robinson, R. Sundararaman, R. Torsi, *MRS Bulletin* **2021**, *46*, 959.

30. G. Liu, S. Rumyantsev, M. A. Bloodgood, T. T. Salguero, M. Shur, A. A. Balandin, *Nano Letters* **2017**, *17*, 377.

31. S. Beyne, K. Croes, I. de Wolf, Z. Tokei, *Journal of Applied Physics* **2016**, *119*, 184302.

32. T. M. Chen, A. M. Yassine, *IEEE Transactions on Electron Devices* **1994**, *41*, 2165.

33. B. Neri, A. Diligenti, P. E. Bagnoli, *IEEE Transactions on Electron Devices* **1987**, *34*, 2317.
34. P. Dutta, P. M. Horn, *Reviews of Modern Physics* **1981**, *53*, 497.
35. W. Yang, Z. Çelik-Butler, *Solid-State Electronics* **1991**, *34*, 911.
36. G. Liu, S. Rumyantsev, M. A. Bloodgood, T. T. Salguero, A. A. Balandin, *Nano Letters* **2018**, *18*, 3630.
37. R. Salgado, A. Mohammadzadeh, F. Kargar, A. Geremew, C. Y. Huang, M. A. Bloodgood, S. Rumyantsev, T. T. Salguero, A. A. Balandin, *Applied Physics Express* **2019**, *12*, 037001.
38. A. K. Geremew, S. Rumyantsev, F. Kargar, B. Debnath, A. Nosek, M. A. Bloodgood, M. Bockrath, T. T. Salguero, R. K. Lake, A. A. Balandin, *ACS Nano* **2019**, *13*, 7231.
39. S. Ghosh, F. Kargar, A. Mohammadzadeh, S. Rumyantsev, A. A. Balandin, *Advanced Electronic Materials* **2021**, *7*, 2100408.
40. S. Ghosh, K. Fu, F. Kargar, S. Rumyantsev, Y. Zhao, A. A. Balandin, *Applied Physics Letters* **2021**, *119*, 243505.
41. S. Ghosh, H. Surdi, F. Kargar, F. A. Koeck, S. Rumyantsev, S. Goodnick, R. J. Nemanich, A. A. Balandin, *Applied Physics Letters* **2022**, *120*, 062103.
42. R. Samnakay, A. A. Balandin, P. Srinivasan, *Solid-State Electronics* **2017**, *135*, 37.
43. P. Gressier, L. Guemas, A. Meerschaut, *Acta Cryst*. 1982, *38*, 2877.
44. P. Gressier, A. Meerschaut, L. Guemas, J. Rouxel, P. Monceau, *Journal of Solid State Chemistry* 1984, *51*, 141.
45. Y. Shi, Q. Li, J. Yu, al -, S. Onishi, M. Jamei, A. Zettl -, S. van Smaalen, E. J. Lam, Journal of Physics: Condensed Matter 2001, 13, 9923.I. Ohana, D. Schmeltzer, D. Shaltiel, Y. Yacoby, A. Mustachi, *Solid State Communications* **1985**, *54*, 747.
46. H. Yi, Z. Huang, W. Shi, L. Min, R. Wu, C. M. Polley, R. Zhang, Y. F. Zhao, L. J. Zhou, J. Adell, X. Gui, W. Xie, M. H. W. Chan, Z. Mao, Z. Wang, W. Wu, C. Z. Chang, *Physical Review Research* **2021**, *3*, 013271.
47. F. Kargar, A. Krayev, M. Wurch, Y. Ghafouri, T. Debnath, D. Wickramaratne, T. T. Salguero, R. K. Lake, L. Bartels, A. A. Balandin, *Nanoscale* **2022**, *14*, 6133.
48. I. Ohana, D. Schmeltzer, D. Shaltiel, Y. Yacoby, A. Mustachi, *Solid State Communications* **1985**, 54, 747.
49. A. Zwick, M. A. Renucci, P. Gressier, A. Meerschaut, *Solid State Communications* **1985**, *56*, 947.

50. S. Sugai, M. Sato, S. Kurihara, *Physical Review B* **1985**, *32*, 6809.

51. T. Sekine, T. Seino, M. Izumi, E. Matsuura, *Solid State Communications* **1985**, *53*, 767.

52. A. K. Geremew, S. Rumyantsev, M. A. Bloodgood, T. T. Salguero, A. A. Balandin, *Nanoscale* **2018**, *10*, 19749.

53. A. Geremew, C. Qian, A. Abelson, S. Rumyantsev, F. Kargar, M. Law, A. A. Balandin, *Nanoscale* **2019**, *11*, 20171.

54. A. A. Balandin, *Nature Nanotechnology* **2013**, *8*, 549.

55. S. M. Shapiro, M. Sato, S. Hoshino, *Journal of Physics C: Solid State Physics* **1986**, *19*, 3049.

56. M. Saint-Paul, P. Monceau, F. Lévy, *Solid State Communications* **1988**, *67*, 581.

57. P. Goli, J. Khan, D. Wickramaratne, R. K. Lake, A. A. Balandin, *Nano Letters* **2012**, *12*, 5941.

58. R. Samnakay, D. Wickramaratne, T. R. Pope, R. K. Lake, T. T. Salguero, A. A. Balandin, Nano Letters **2015**, *15*, 2965.

59. C. Tournier-Colletta, L. Moreschini, G. Autès, S. Moser, A. Crepaldi, H. Berger, A. L. Walter, K. S. Kim, A. Bostwick, P. Monceau, E. Rotenberg, O. v. Yazyev, M. Grioni, *Physical Review Letters* **2013**, *110*, 236401.

60. Z. Huang, H. Yi, L. Min, Z. Mao, C. Z. Chang, W. Wu, *Physical Review B* **2021**, *104*, 205138.

61. J. Wu, Q. Gu, B. S. Guiton, N. P. de Leon, L. Ouyang, H. Park, *Nano Letters* **2006**, *6*, 2313.

62. H. M. Lefcochilos-Fogelquist, O. R. Albertini, A. Y. Liu, *Physical Review B* **2019**, *99*, 174113.

63. M. K. Lin, J. A. Hlevyack, P. Chen, R. Y. Liu, S. K. Mo, T. C. Chiang, *Physical Review Letters* **2020**, *125*, 176405.

64. D. J. Ulness, A. C. Albrecht, *Physical Review A* **1996**, *53*, 1081.

65. T. Grasser, *Springer Nature* **2020**.

66. S. L. Rumyantsev, C. Jiang, R. Samnakay, M. S. Shur, A. A. Balandin, *IEEE Electron Device Letters* **2015**, *36*, 517.

67. M. A. Stolyarov, G. Liu, S. L. Rumyantsev, M. Shur, A. A. Balandin, *Applied Physics Letters* **2015**, *107*, 023106.

68. S. Y. Xu, N. Alidoust, I. Belopolski, Z. Yuan, G. Bian, T. R. Chang, H. Zheng, V. N. Strocov, D. S. Sanchez, G. Chang, C. Zhang, D. Mou, Y. Wu, L. Huang, C. C. Lee, S. M. Huang, B. Wang, A. Bansil, H. T. Jeng, T. Neupert, A. Kaminski, H. Lin, S. Jia, M. Z. Hasan, *Nature Physics* **2015**, *11*, 748.
69. M. Z. Hasan, G. Chang, I. Belopolski, G. Bian, S. Y. Xu, J. X. Yin, *Nature Reviews Materials* **2021**, *6*, 784.
70. L. K. J. Vandamme, *IEEE Transactions on Electron Devices* **1994**, *41*, 2176.
71. J. Xu, D. Abbott, Y. Dai, *Microelectronics Reliability* **2000**, *40*, 171.
72. Z. Gingl, C. Pennetta, L. B. Kiss, L. Reggiani, *Semiconductor Science and Technology* **1996**, *11*, 1770.
73. A. A. Balandin, *American Scientific Publishers* **2002**, 258.